\documentclass[twocolumn,final,times,sort&compress]{elsarticle}
\usepackage{graphicx,amssymb,amsmath}
\usepackage{caption}
\usepackage{tikz}
\usepackage{natbib}
\usepackage{soul}
\usepackage{color}
\journal{Physica B}

\begin{document}

\begin{frontmatter}

\title{The effect of relativity on stability of Copernicium phases, their
electronic structure and mechanical properties\tnoteref{grant}}
\tnotetext[grant]{D. L. acknowledges support by The Ministry of Education, Youth and Sports from the Large Infrastructures for Research, Experimental Development 
and Innovations project „IT4Innovations National Supercomputing Center – LM2015070“ and by the grant No. 17-27790S of the Grant Agency 
of the Czech Republic and by The Ministry of Education, Youth and Sports from the National Programme of Sustainability (NPU II) project 
"IT4Innovations excellence in science - LQ1602" and  Mobility grant No. 8X17046 and Student Grant Competitions of VSB-TU Ostrava (SP2017/184).
 H.C. acknowledges support by the Slovak Research and Development Agency (APVV) under Grant No. DS-2016-0046. The financial support provided by the ERDF EU 
Grant under the contract No. ITMS 26110230061 is also gratefully acknowledged. This work was supported by the European Regional Development Fund in the  IT4Innovations national supercomputing center$-$path to exascale project, project number CZ.02.1.01/0.0/0.0/16\_013/0001791  within the Operational Programme Research, Development and Education.} 
\author[UEFSAV]{Hana \v Cen\v carikov\'a} 
\author[IT4I]{Dominik Legut\corref{coraut}}
\cortext[coraut]{Corresponding author}
\ead{dominik.legut@vsb.cz}
\address[UEFSAV]{Institute  of  Experimental  Physics,  Slovak   Academy   of Sciences, Watsonova 47, 040 01 Ko\v {s}ice, Slovakia}
\address[IT4I]{IT4Innovations Center, VSB – Technical University of Ostrava, 17.listopadu 15, 708 33 Ostrava-Poruba, Czech Republic}

\begin{abstract}
The phase stability of the various crystalline structures of the super-heavy element Copernicium  was determined based on the first-principles calculations with 
different levels of the relativistic effects. We utilized the Darwin term, mass-velocity, and spin-orbit interaction with the single electron framework of the density 
functional theory while treating the exchange and correlation effects using local density approximations. It is found that the spin-orbit coupling is the key component 
to stabilize the body-centered cubic ($bcc$) structure over the hexagonal closed packed ($hcp$) structure, which is in accord 
with Sol. Stat. Comm. 152 (2012) 530, but in contrast to Sol. Stat. Comm. 201 (2015) 
88, Angew. Chem. 46 (2007) 1663, Handbook of Elemental Solids Z=104-112 (Springer 2015). It seems that the main role here is the 
correct description of the semi-core relativistic $6p_{1/2}$ orbitals. The all other investigated structures, {\it i.e.} face-centered cubic 
($fcc$), simple cubic ($sc$) as well as rhombohedral ($rh$) structures are higher in energy. The criteria of mechanical stability were 
investigated based on the calculated elastic constants, identifying the phase instability of $fcc$ and $rh$ structures, but surprisingly confirm the 
stability of the energetically higher $sc$ structure. In addition, the pressure-induced structural transition between two stable $sc$ and $bcc$ phases has 
been detected. The ground-state $bcc$ structure exhibits the highest elastic anisotropy from single elements of the Periodic table. At last, we support 
the experimental findings that Copernicium is a metal.  

\end{abstract}


\end{frontmatter}

\section{Introduction}

Element with 112 protons in nuclei, known as Copernicium (Cn), belongs to the group of transactinides and together with other fourteen known chemical 
elements completes the 
12$^{th}$ group of the Periodic table. These elements, also known as super-heavy elements (SHEs), were firstly detected accidentally at the sixth decade of twenty 
century in the 
experiment realized by Polikanov et al.~\cite{Polikanov}, in which the new element $Rf$ (with the atomic number $Z=104$) was identified as a ''by-product'' of 
spontaneuosly fissioning isomers. The discovery of this new heavy element started up an enormous effort of researchers to identify in nature or to synthesize artificially 
other atoms with a higher number of protons in their nucleus, expecting unconventional electronic, geometric, and magnetic structure properties. Unfortunately, their 
effort to identify aforementioned elements in nature remained unsuccessful however the advances in experimental capabilities in combination with a rapid progress in 
computer science technique allows today to synthesize SHE atoms with atomic numbers up to $Z=117,118$ \cite{Khuyagbaatar14,Oganessian06}. The  very 
short 
half-life of induced SHE atoms, usually only a few milliseconds~\cite{Morss,Kaldor,Schadel}, leads to the consecutive suppression of initial interest in SHE. The 
direct consequences of their short half-life are reflected  in a low production rates of experiment, where {\it e.g.} generates only  one atom during some days~\cite{Hofmann} 
and also in production of very unstable elements without their next application. Even the later prediction of the so-called ``island of stability'' which suggests that 
there are several stable very heavy elements~\cite{Myers}, re-started the activities in this research area. Just the technical demandingness of 
synthesis shows on the 
necessity of precise theoretical support. 

Among all SHEs, the special attention has been devoted to the element \#112, Copernicium, for the fact that just this element 
(similarly as the latest element ``Og'' with the $Z=118$) has an electron structure with all closed shells, indicating a similarity to the elements Hg and Rn of the 
$6^{th}$ level of the  Periodic table. Recently, a large debate has opened for the ground-state structure of Cn, {\it i.e.} face-centered cubic the
($fcc$)~\cite{Dimitris}, 
the hexagonal close pack ($hcp$)~\cite{Gaston,Atta}, or the body centered cubic ($bcc$)~\cite{Zaoui}. Another controversy is 
whether the Cn is of metallic character or not. The 
band gap was predicted in Refs.~\cite{Dimitris,Gaston,Atta}. However, recent experimental study shows that the Cn is rather a volatile 
metal~\cite{Yakushev14} as predicted already in 1975 by Pitzer~\cite{Pitzer75}.  It should be mentioned, that such ambiguity has not been 
observed in remaining 
investigated SHEs for which a pure metallic 
behaviour has been detected~\cite{Dimitris,Gyanchandani}. The discrepancies in manifold results originate from the diversity of used computational 
methodologies, {\it 
i.e.} at which level the relativistic effects are taken into the account. 
To bring the light into this problem, in this paper we have investigated the energetics and mechanical stability of the Cn atom in five basic Bravais lattice structures, 
namely  the $bcc$, the $fcc$, the $hcp$, the simple cubic ($sc$) and finally the 
rhombohedral ($rh$) structure using the framework of density functional 
theory calculations. 
To treat the relativistic effects  and to explain the discrepancy among previous results we adopted following approaches; the non-relativistic (NR), the 
scalar-relativistic (SR), the scalar-relativistic calculation with the 
inclusion of the spin-orbit interaction (SOC) and the SOC calculations with additional basis functions (local orbitals) for the low lying $6p$ semi-core 
states (RLO).  
 We suppose, that our complex analysis allows us to find a definite answer on the question about the crystallography preference as well 
as 
conduction properties of the super-heavy element Cn. 

The paper is organized as following; in Section~\ref{Methodology} we briefly describe used methodology with computational details, involving the conditions of 
mechanical stability for all five investigated structures. Subsequently, the main results based on the calculated phase energetics and elastic constants are present in 
Section~\ref{Results}. Finally, the most significant results   are summarized together with future outlooks in the Section~\ref{Conclusion}.

\section{Computational details}
\label{Methodology}
\begin{figure}[ht!]
\begin{center}
{\includegraphics[width=0.5\textwidth,height=0.25\textheight,trim={3cm 9.3cm 3.5cm 10.9cm},clip]{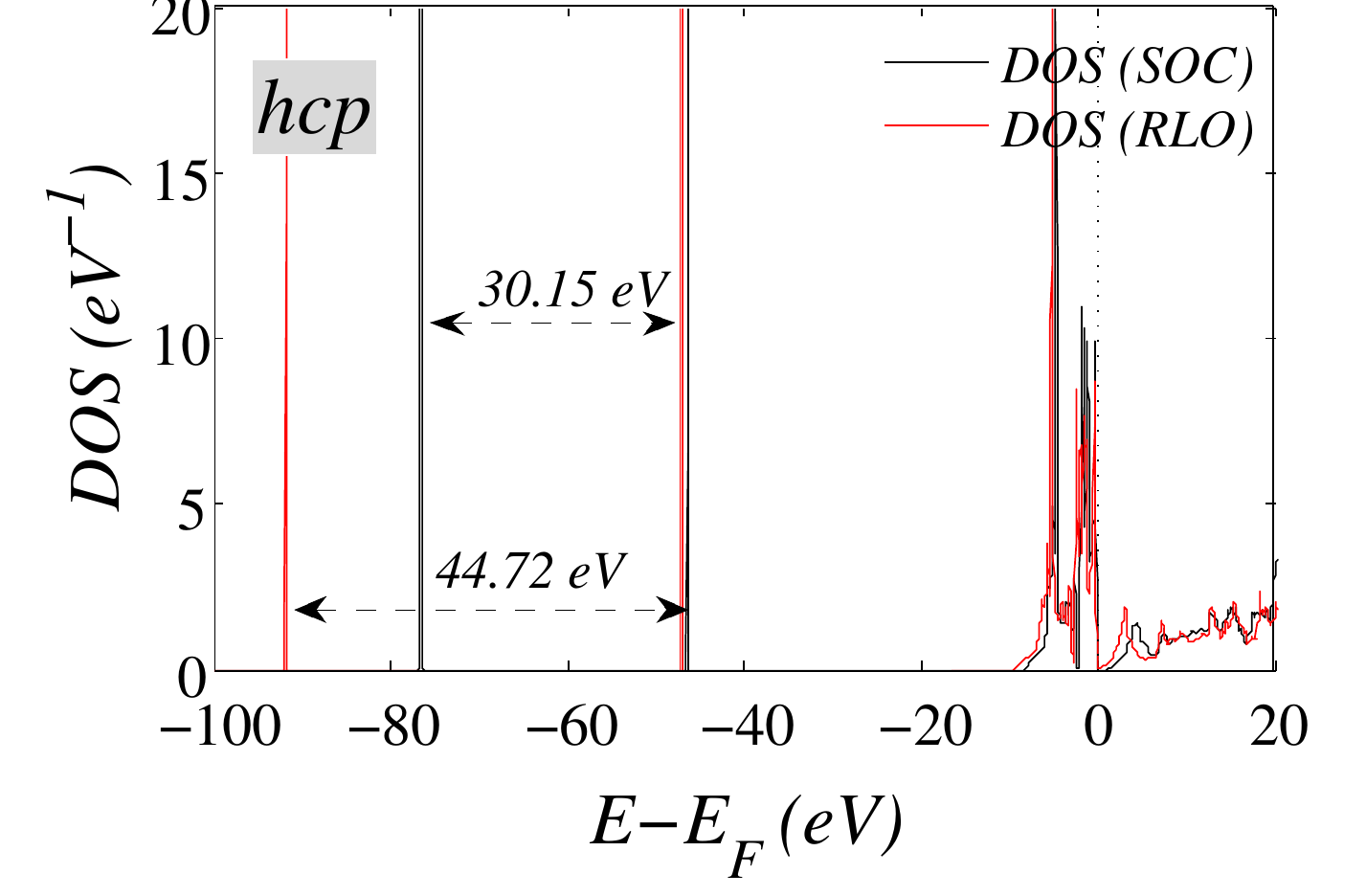}}\\
{\includegraphics[width=0.5\textwidth,height=0.25\textheight,trim={3cm 9.3cm 3.5cm 10.3cm},clip]{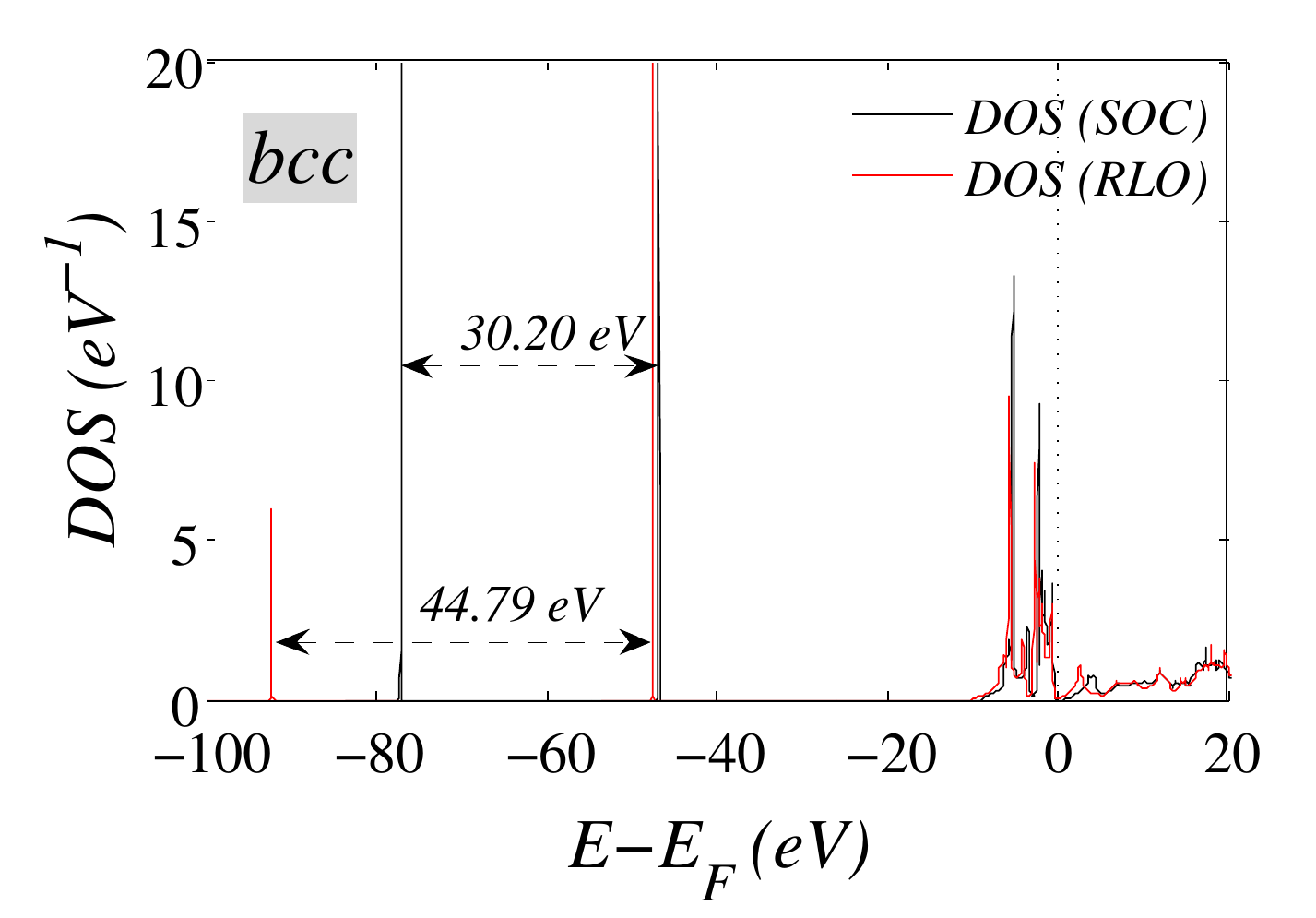}}\\
{\includegraphics[width=0.45\textwidth,height=0.24\textheight,trim={3cm 9.5cm 2.7cm 10.5cm},clip]{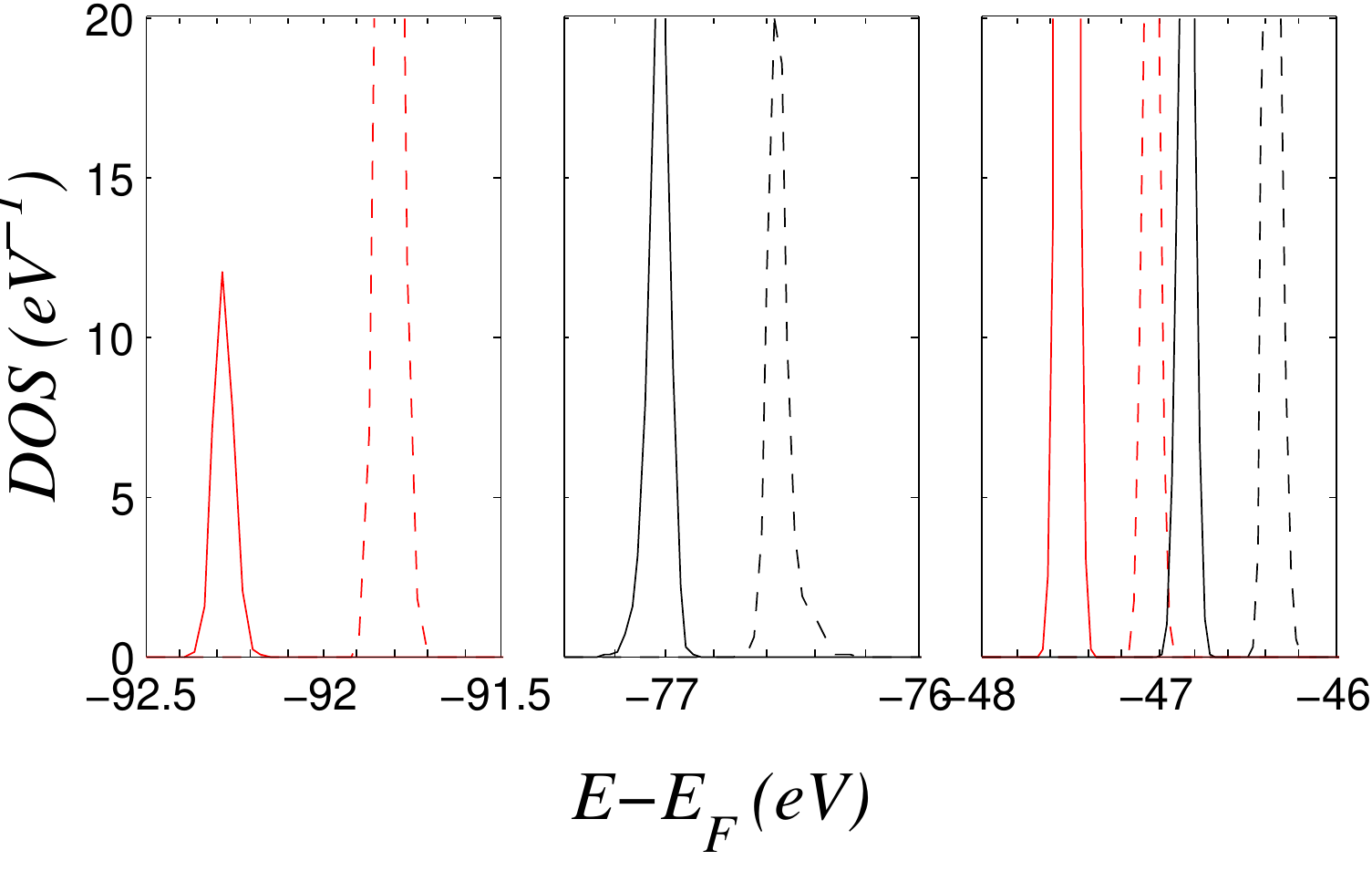}}\\
\caption{ Total electron density of states per atom for the $hcp$ (upper panel) and $bcc$ (middle panel) structure of the \#112 
Cn calculated at the SOC and RLO level. For the RLO, the splitting of the $6p_{1/2}$ and $6p_{3/2}$ is larger than in SOC case. In addition please note the far position 
for the $6p_{1/2}$ state from Fermi level for the RLO than 
for the SOC calculations. For a better visualization the details of 6$p$ states are done  for both lattices (lower panel). The $hcp$ structure corresponds to the 
dashed lines while the $bcc$ one to the solid lines. The  SOC  (the black lines) and the RLO (the red lines) calculations.}
\label{fig2}
\end{center}
\end{figure}

A systematic study has been performed using the full potential linearized augmented plane wave method implemented in the WIEN2k code~\cite{Blaha} assuming the exchange 
and correlation effects through the local density approximation (LDA)~\cite{Perdew}. The basis function has been expanded up to $R_{MT}K_{MAX}$=15, where $R_{MT}$ is the 
muffin-tin radius of 2.40 bohr and  $K_{MAX}$ is the maximum modulus for the reciprocal lattice vectors. The maximum value of partial waves inside the atomic sphere has  
been set as $l_{max}$=12. Highly accurate Brillouin zone  integrations are performed using the standard special k-points technique of Monkhorst and Pack 
(MP)~\cite{Monkhorst} with a $19\times19\times19$ MP mesh and all self-consistent calculations have been performed with the energy convergence criteria better than $1E-6$ 
Ry/atom. The core electrons, here [Xe]$4f^{14},5d^{10},6s^2$ are treated by Dirac approach~\cite{Desclaux},  the valence electrons $6p^6,5f^{14},6d^{10},7s^2$ are treated 
by the  {so-called scalar-relativistic (SR) calculations (solving Schroedinger equations). The energy cut-off between the core and valence region has been set as -9 Ry. 
The perturbation of the spin-orbit interaction then added on top within the second-variation method~\cite{Koelling} (SOC calculation) with the cut-off energy 
$E_{cut}$=6 Ry. However, this approach still suffers from none-relativistic basis functions and was shown not to be sufficient to address correct description of the 
semi-core $6p$-states in $fcc$ Th~\cite{Kunes} and in $fcc$ Pb~\cite{Singh}. For the elements with high atomic number where large SOC is expected, the $p_{1/2}$ and 
$s_{1/2}$ radial wave functions have finite amplitudes at the position of the point like nucleus~\cite{Autschbach,Cremer}.
Therefore, the additional local orbitals for the $p$ semi$-$core states were added, where the radial part of the $p_{1/2}$ basis is obtained of the Dirac equations for 
$l=1$ and $j=1/2$ (RLO calculation)~\cite{Kunes,Singh}. In this approach, the non-zero radial wave function of low-lying $p_{1/2}$ states 
extends the basis sets, which allow better convergence of the calculations, {\it e.g.} $fcc$ Th~\cite{Kunes} as well as the correct splitting and stabilization of the 
$p$ semi-core states~\cite{Kunes,Singh}. For more details about the treatment of the relativistic effects, see Ref.~\cite{Strange, Autschbach,Cremer} and references 
therein. To analyze the mechanical stability of Cn, the second order  elastic constants $c_{ij}$ of all five lattice structures were determined from the dependence of 
total energy vs. corresponding deformations. For this purpose, the packages ``ELAST`` and ''IRELAST``~\cite{Charpin, Jamal}, as implemented in the WIEN2k code have been  
used. Subsequently, the evaluation of the selected $c_{ij}$ has been performed through the  fourth order fit. To judge the mechanical stability, we have the following 
criteria~\cite{Mouhat14}, which must hold for the stable structure:
\\
For cubic structures ({\it sc, fcc, bcc}): 
\begin{align}
&c_{11}-c_{12}>0,\;\;\; c_{11}+2c_{12}>0,\;\;\; c_{44}>0.
\label{eq1}
\end{align}
For the {\it hcp} structure:
\begin{align} 
&c_{11}-|c_{12}|>0,\;\; c_{33}(c_{11}+c_{12})-2c_{12}^2>0,\;\;\; c_{44}>0,
\label{eq2}
\end{align}
and finally, for the {\it rh} structure: 
\begin{align} 
&c_{11}-|c_{12}|>0,\;\;\; c_{33}(c_{11}+c_{12})-2c_{13}^2>0,\nonumber\\
&c_{44}(c_{11}-c_{12})-2c_{14}^2>0,\;\;\;c_{44}>0.
\label{eq3}
\end{align}
For the cubic stability, the first two conditions could be reduced 
to the tetragonal elastic constant $c'=\frac{c_{11}-c_{12}}{2}$, which must be positive if the structure is stable.
 In addition, we define the elastic anisotropy factor~\cite{Zener}  for all cubic structures as $A=c_{44}/c'$, as a system response 
to the type of applied deformation. In general, if the $A$ equals unity, the solid behaves as an isotropic one~\cite{Grimvall}. \\
\section{Results and discussion}
\label{Results}

Let us first discuss the results along with the increase of the relativistic effects on the phase stability of the five different 
structures of Cn. We determined the equilibrium volume and calculated the corresponding total energy  with respect to the ground state at each level of the 
inclusion of the relativistic effects, {\it i.e.} the NR calculation, SR calculation, SOC calculation as well as RLO 
one, see  Tables~\ref{taba}-\ref{tabd} in Appendix. As is evident for the NR, SR, 
and SOC analyses, the ground-state structure is the $hcp$ one, which corresponds to the previous results obtained in the SR or SOC limit~\cite{Gaston, 
Atta}. Our 
calculations in addition showed that also the different results presented in work}~\cite{Dimitris} are numerically correct, not physically, since 
an incomplete set of lattice structures has been assumed. 
 With increase of relativistic effects, the energy difference between the $hcp$, and $bcc$ structure decreases and finally 
the $bcc$ structure becomes the ground state which is now preferred by ca. 1mRy/atom. 
We suppose that the origin of this significant change in the structure preference arises from the  basis extension, including also the non-zero radial wave 
function in the RLO calculation. To verify this conjecture, we analyze the $p$- and $d$-density of states (DOS) per atom in the SOC and RLO case for the $bcc$ as well 
as $hcp$ structures, see Fig.~\ref{fig2}. 
 As one can see, the splitting of the $6p_{1/2}$ and $6p_{3/2}$ is larger when the relativistic 
basis for the $p_{1/2}$ is taken into account than in the SOC calculations, similar to the case of $fcc$ Th~\cite{Kunes}. 
The comparison of $p$-DOSs/atom of $hcp$ and $bcc$ structures exhibits different position of $p_{1/2}$ level. In the $bcc$ structure, $p_{1/2}$ 
states with 
respect to the Fermi level lie deeper in the energy than its $hcp$ counterpart, while the position of $p_{3/2}$ states remains almost unchanged for both phases. This has 
a stabilization effect for the $bcc$ structure over the $hcp$ one. 

\begin{figure}[h!]
\begin{center}
{\includegraphics[width=0.45\textwidth,trim={2.5cm 9cm 3.5cm 9.5cm},clip]{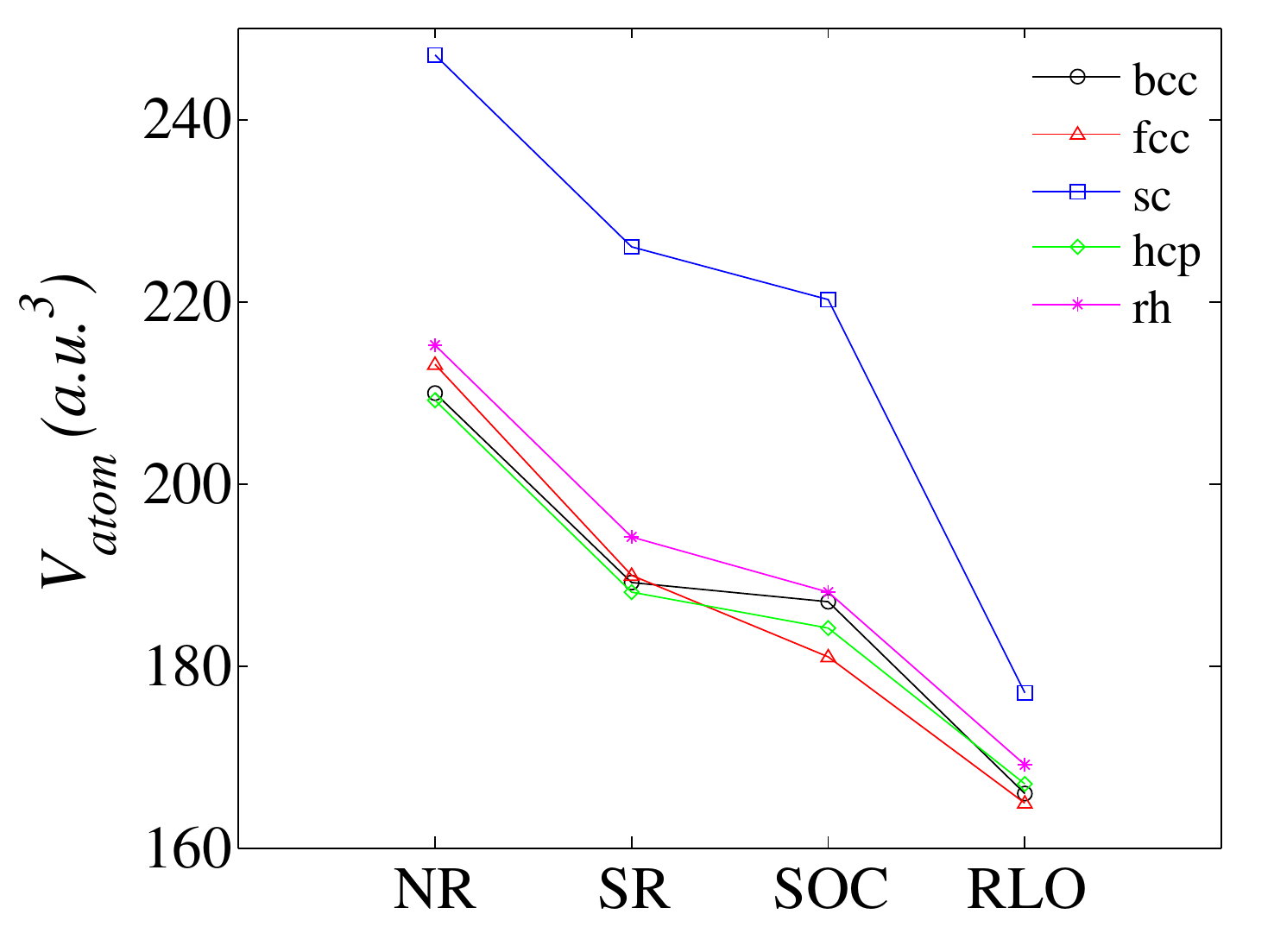}}
\caption{The evolution of equilibrium volume ($V_{atom}$) depends on the used approximations (NR, SR, SOC and RLO) for all five investigated lattice 
structures.}  \label{fig1a}
\end{center}
\end{figure}

\begin{figure}[h!]
\begin{center}
{\includegraphics[width=0.45\textwidth,trim={1cm 7cm 1.5cm 8cm},clip]{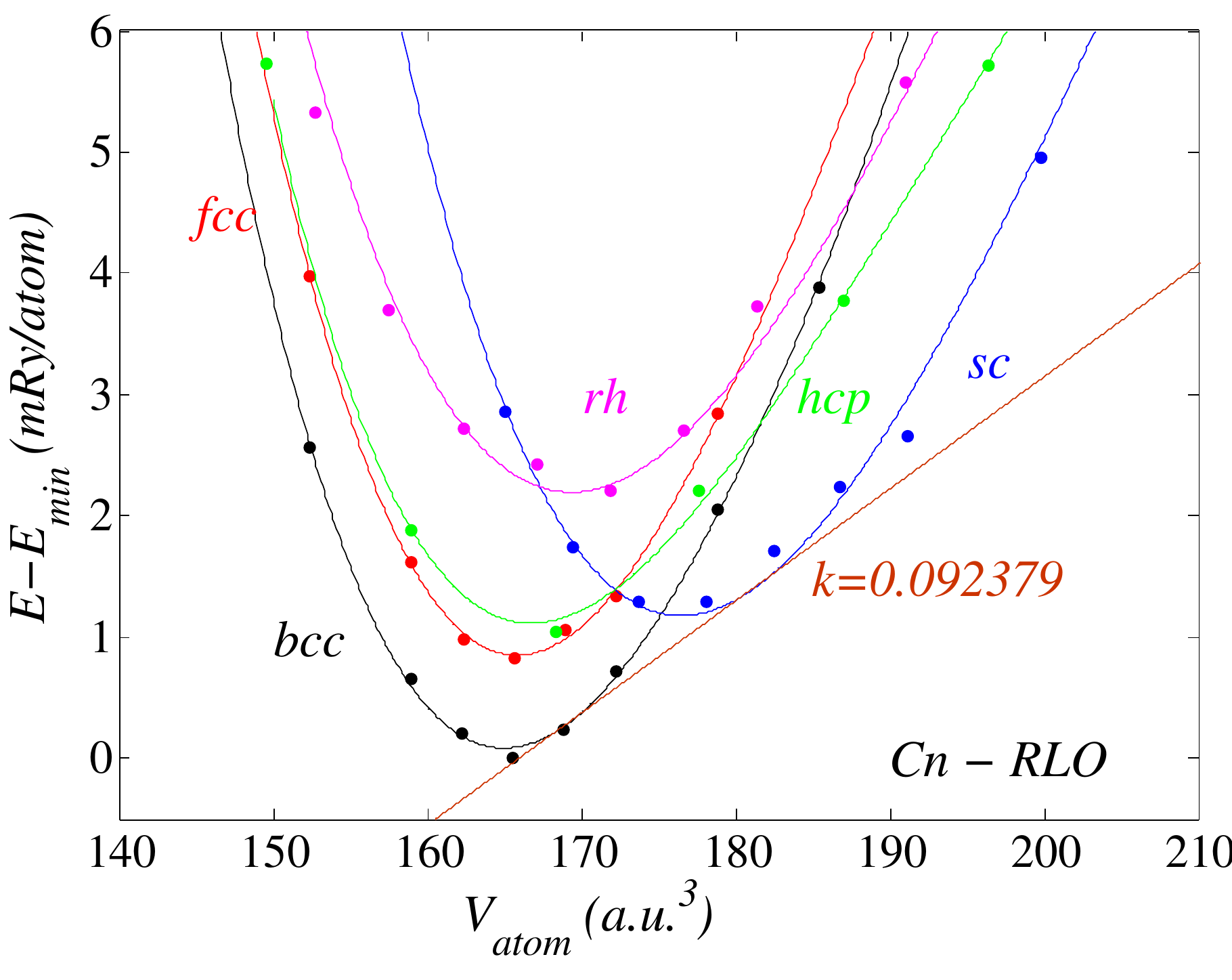}}
\caption{Total energy as a function of volume for the \#112 Cn element calcualated at the RLO level. 
The orange line in figure denotes a common tangent with a slope $k=0.092379$ of $sc$ as well as $bcc$ curves. }
\label{fig1}
\end{center}
\end{figure}

The ``increase'' of relativity is similar to the pressure,  where the volumes of all structures decrease along NR$\rightarrow$SR$\rightarrow$SOC$\rightarrow$RLO 
 line. This is nicely visible for the $sc$ structure (see Fig.~\ref{fig1a}), 
which has 
a filling factor of 52\%. In addition, the increase of relativity significantly influences an energy relation among all investigated structures. For example, 
the difference between the structure with the lowest energy ({\it hcp}) and the structure with the highest energy ({\it sc}) at the NR calculations is around  19mRy/atom, 
while at the  RLO level it is only 1.25mRy/atom, see Table~\ref{tabd}. 
Similarly from the NR towards to RLO calculations, the volume of the $bcc$ structure (filling 68\%) decreases to the 77\%, whereas the $hcp$ 
(filling 
74\%) one only to 80\%.  This also implies a more strong bonding for the $bcc$ structure over the $hcp$ one. The second  interesting 
observation, derived from the energetic analysis of RLO case is the fact, that  there exists a common  tangent between the $sc$ and $bcc$ structure, 
 which indicates on the possible pressure-induced structure transformation from the $sc$ to $bcc$ structure at the 
pressure of 1.359 GPa (see Fig.~\ref{fig1}). 
This transition should be possible if and only if both structures are stable. For this reason we calculated the elastic constants ($c_{ij}$) in the RLO limit, but also in 
 the NR, SR as well as SOC limit, with the aim to understand the evolution of stable structure under the influence of relativistic effects in context of previous results 
(see right column of Tables~\ref{taba}-\ref{tabd}).  Concerning the mechanical stability criteria, (Section~\ref{Methodology}) we found that for the NR case  besides the 
$hcp$ structure also the $fcc$ structure is stable and in addition both structures are very close to the volume as well as to the energy in the equilibrium state. For 
the SR calculations only the $fcc$, $rh$ and $hcp$  lattices are stable and only the $fcc$ and $hcp$ one at SOC level. Surprisingly, our 
$E$-$V$ analyses indicate that detected 
stable states (in the NR as well as SR calculations) are realizable under the modulation of external pressure. For the most interesting RLO calculations it was found 
that 
both mentioned structures, $bcc$ and $sc$, are mechanically stable, and thus the pressure-driven structural transition can be possible. Besides them,  also the $hcp$ 
structure satisfies the stability conditions Eq.~(\ref{eq2}), but as illustrates Fig.~\ref{fig1} its realization seems to be impossible. The detailed 
analysis of mechanical 
stability showed on the another fascinating feature of Cn element, and namely, that the stable $bcc$ structure (in the RLO case) has very low $c'=4$GPa, {\it i.e.} 
resistance to the tetragonal deformation and therefore  has very high elastic anisotropy $A=9.75$. Such value is exceeding the one of pure Li ($A=8.524$) and is the 
highest among the single element of the Periodic table and three orders of magnitude of the other bound $sc$ Po~\cite{leg07}.  

\begin{figure}[h!]
\begin{center}
\vspace*{-0.4cm}{\it bcc}\\
{\includegraphics[width=0.4\textwidth,height=0.29\textheight,trim={0cm 0cm 0cm 0cm},clip]{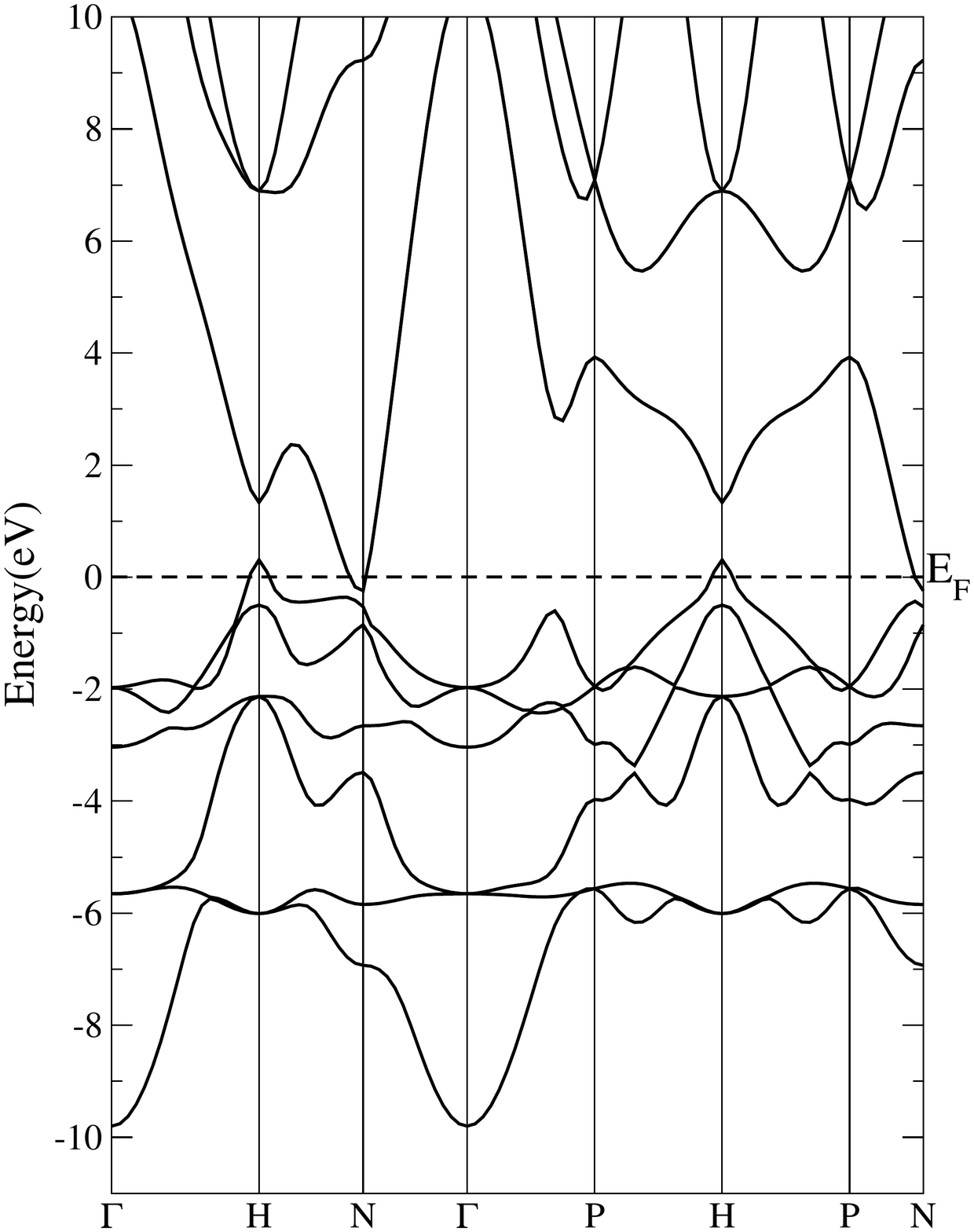}}\\\vspace*{-0.4cm}{\it sc}\\
{\includegraphics[width=0.4\textwidth,height=0.29\textheight,trim={0cm 0cm 0cm 0cm},clip]{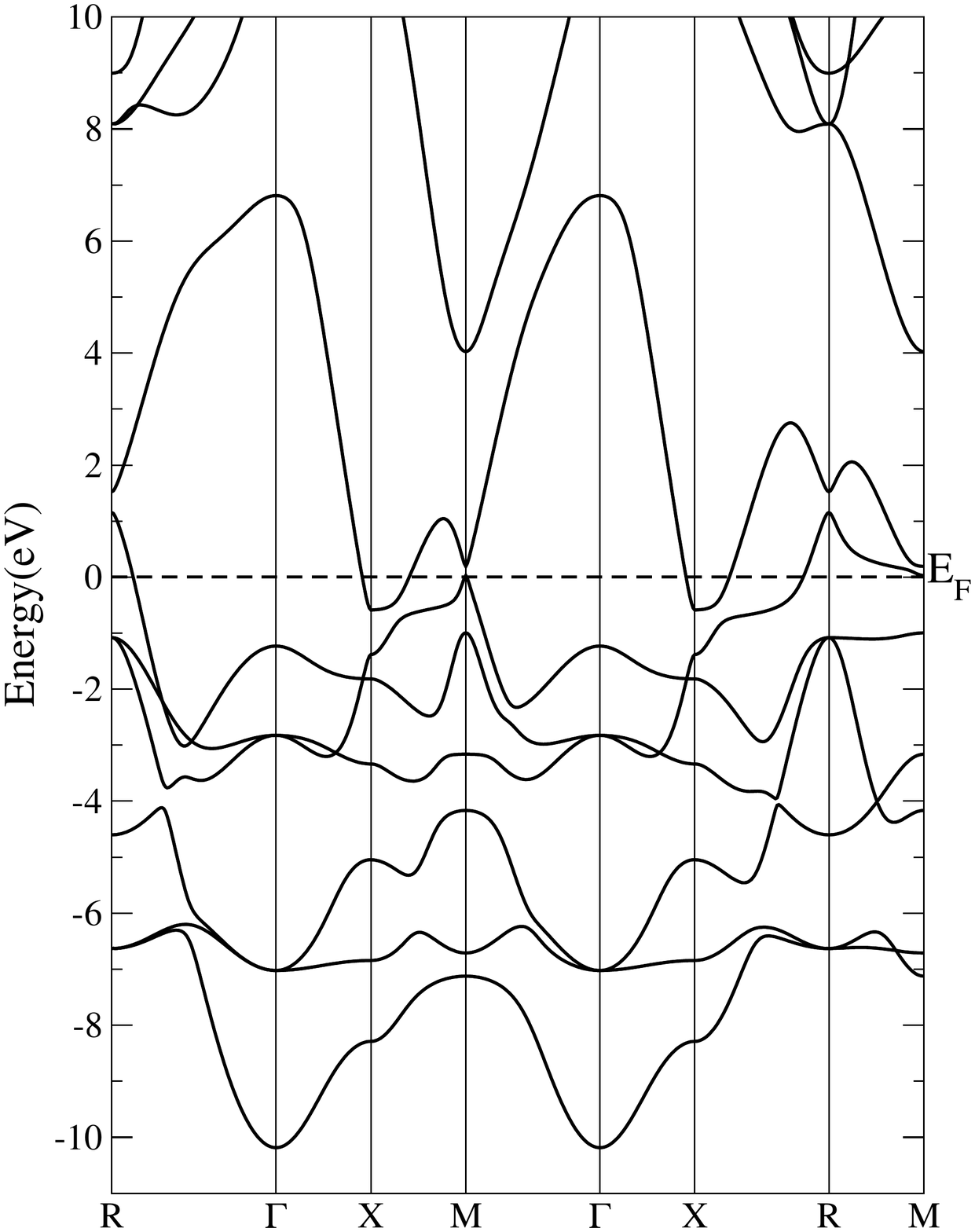}}\\\vspace*{-0.4cm}{\it hcp}\\
{\includegraphics[width=0.4\textwidth,height=0.29\textheight,trim={0cm 0cm 0cm 0cm},clip]{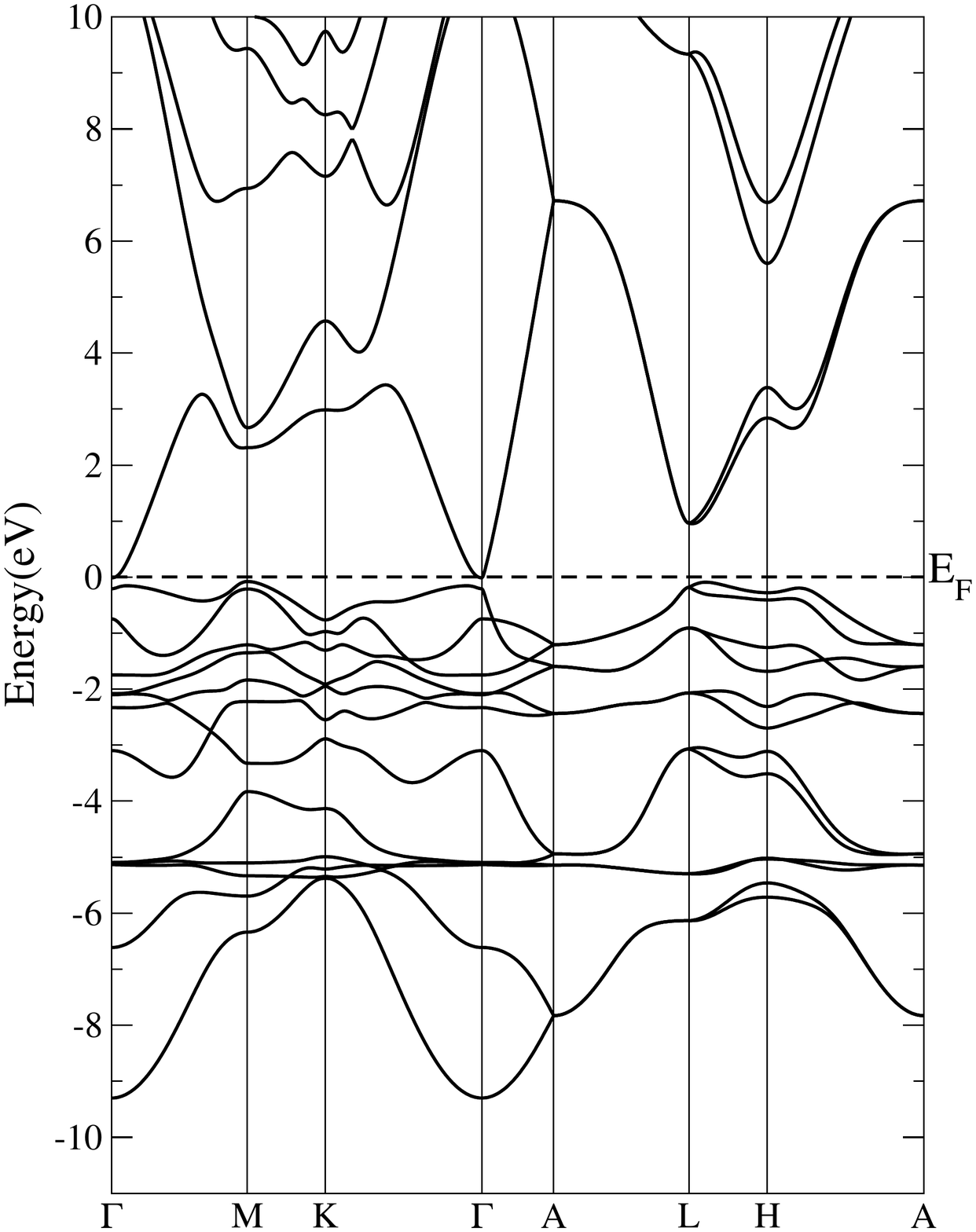}}
\caption{ The band structure of the $bcc$ structure (upper panel), the  $sc$ structure (middle panel), and the $hcp$ structure (lower panel) of the Cn 
calculated at RLO level.}  
\label{fig3}
\end{center}
\end{figure}
At last but not least, we examined the band structure of energetically favored $bcc$ structure in the RLO limit with the aim to confirm or confute metallic or insulating 
behavior of the Cn element. Obtained results are present in Fig.~\ref{fig3} (the upper panel). Evidently, the Fermi energy intersects the valence orbital 
(the $s$ orbital) as well as the conducting orbital (the $p$ orbital) and thus the Cn element in the $bcc$ structure is really metallic as was found 
experimentally~\cite{Yakushev14}. In addition, we have found that the second stable $sc$ phase  which can be reached under the pressure 
influence, exhibits a metallic behavior, contrary to the last stable $hcp$ structure with an indirect gap of 27.2meV (nicely visible from Fig.~\ref{fig3}, 
the middle and lower part of figure). 
\section{Conclusion}
\label{Conclusion}

In the presented paper we have explained the discrepancy among previous partial results about the ground-state structure of the super-heavy element Cn as well as its 
conducting properties.  We have shown that the key factor that affects the stability of the Cn is the correct description ({\it energy position}) of the $6p_{1/2}$ 
states 
that could be taken into account only if the relativistic local orbitals are taken into the basis. In addition, we have found that the relativistic effects causing the 
electronic states closer to nuclei to contract act as ``pressure'' results to the energy preference of the $bcc$ structure instead of the  $hcp$ one. Besides, we have 
determined that the Cn element can exist at two possible stable states, namely, at the $bcc$ structure (with the lowest enthalphy) as well as the $sc$ structure, if the 
volume of the phase could be expanded (negative pressure).  An exhaustive analyse of elastic constants showed that the Cn element at the $bcc$ structure has very low 
resistance to the tetragonal deformation ($c'$) and therefore exhibits the largest elastic anisotropy $A=9.75$ among of simple elements of the Periodic table 
exceeding the one of the Li $A=8.52$ and become the opposite bound of $sc$ Po ($A=0.06$)~\cite{leg07}. Finally, we have found that the Cn element is a metal at both 
stable states in accord with recent theoretical prediction~\cite{Yakushev14}. 

\section{Appendix}
The lattice constants, the total energy difference  to the ground-state ($\Delta E$), equilibrium volume ($V_{at}$), bulk modulus ($B$), and elastic constants 
($c_{ij}$) are shown for NR calculations (Table \ref{taba}), SR calculations (Table \ref{tabb}), SOC calculation (Table \ref{tabc}), and finally for the RLO calculations 
(Table \ref{tabd}).  

\begin{table}[h!]
\begin{center}
\resizebox{0.5\textwidth}{!} {
\begin{tabular}{c|c|c|c|c|c|c}
\hline
& $a$ [a.u.] & $c$ [a.u.] &  $\Delta E$[mRy/at.] & $V_{at.}$ [a.u.$^3$] & $B$ [GPa] & $c_{ij}$ [GPa]\\
\hline\hline
fcc (S)  & 9.4356 & &  1.1  & $\begin{array}{c}210\\A=0.013\end{array}$ & 47  & $\begin{array}{c}c_{11}=67 \\c_{12}= 37 \\c_{44}=0.2 
\end{array}$\\
\hline
bcc (U)  & 7.5190 & & 5.9 & $\begin{array}{c}213\\A=-0.17\end{array}$ & 45  &$\begin{array}{c}c_{11}= 29\\c_{12}=53 \\c_{44}= 2 (2)\end{array}$\\
\hline
sc (U)  & 6.2762 & & 18.7  & $\begin{array}{c}247\\A=-0.26\end{array}$ & 33 & $\begin{array}{c}c_{11}=64 \\c_{12}=17 \\c_{44}=-6\end{array}$\\
\hline
hcp (S)  & 6.5184 & 11.3365 & 0 & $\begin{array}{c}209\\\end{array}$&  48 & $\begin{array} {c}c_{11}=89\\c_{12}=36\\c_{13}=35\\c_{33}=54\\c_{55}=24\end{array}$\\
\hline
rh (U)  & 7.5649 & 86.8$^o$ & 6.8 & $\begin{array}{c}215\\\end{array}$& 50  
&$\begin{array}{c}c_{11}=58\\c_{12}=43\\c_{13}=47\\c_{14}=1\\c_{33}=38\\c_{44}=-7\end{array}$\\
\hline
\end{tabular}
}
\caption{The lattice constants, the energy difference $\Delta E=E-E_{min}$ with respect to the lowest phase, volume and elastic anisotropy $A$, bulk modulus, and elastic 
constants 
for the $bcc$, $fcc$, $sc$, $hcp$, and $rh$ structures for the non-relativistic (NR) calculations. Abbreviations (S) and (U) denote the stable and unstable phase in 
order,  using the stability criteria Eqs.~(\ref{eq1})-(\ref{eq3}).} 
\label{taba}
\end{center}
\end{table}
\vspace*{-0.9cm}
\begin{table}[h!]
\begin{center}
\resizebox{0.5\textwidth}{!} {
\begin{tabular}{c|c|c|c|c|c|c}
\hline
& $a$ [a.u.] & $c$ [a.u.] &  $\Delta E$[mRy/at.] & $V_{at.}$ [a.u.$^3$] & $B$ [GPa] & $c_{ij}$ [GPa]\\
\hline\hline
fcc (S) & 9.1080 & & 0.1  & $\begin{array}{c}189\\A=3.5\end{array}$ & 35  & $\begin{array}{c}c_{11}=39.5\\c_{12}= 
32.0\\c_{44}=13 \end{array}$\\
\hline
bcc (U) & 7.2443 & & 1.0 & $\begin{array}{c}190\\A=-17\end{array}$ & 34 &$\begin{array}{c}c_{11}= 33\\c_{12}=35\\c_{44}= 17 \end{array}$\\
\hline
sc (U) & 6.0901 & & 9.9 & $\begin{array}{c}226\\A=-0.07\end{array}$ & 22 & $\begin{array}{c}c_{11}=57\\c_{12}=4 \\c_{44}=-2\end{array}$\\
\hline
hcp (S) & 6.44297 & 10.4847 & 0 & $\begin{array}{c}188\\\end{array}$&  35 & $\begin{array}{c}c_{11}=49\\c_{12}=29\\c_{13}=27\\c_{33}=52\\c_{55}=7\end{array}$\\
\hline
rh (S) & 7.2930 & 88.8$^o$ & 2.1 & $\begin{array}{c}194\\\end{array}$& 32  
&$\begin{array}{c}c_{11}=49\\c_{12}=27\\c_{13}=20\\c_{14}=0.02\\c_{33}=51\\c_{44}=4\end{array}$\\
\hline
\end{tabular}
}
\caption{The same as in Table \ref{taba} but for the SR calculations, {\it i.e.} Darwin and mass-velocity term are included.} 
\label{tabb}
\end{center}
\end{table}
\begin{table}[h!]
\begin{center}
\resizebox{0.5\textwidth}{!} {
\begin{tabular}{c|c|c|c|c|c|c}
\hline
& $a$ [a.u.] & $c$ [a.u.] &  $\Delta E$[mRy/at.] & $V_{at.}$ [a.u.$^3$] & $B$ [GPa] & $c_{ij}$ [GPa]\\
\hline\hline
fcc (S) & 9.0726 & & 0.40  & $\begin{array}{c}187\\A=3.8\end{array}$ & 38 & $\begin{array}{c}c_{11}=40\\c_{12}=32 
\\c_{44}=15 \end{array}$\\
\hline
bcc (U) & 7.1243 & & 0.69 &  $\begin{array}{c}181\\A=-4\end{array}$ & 47  &$\begin{array}{c}c_{11}=28 \\c_{12}=57 \\c_{44}=58 \end{array}$\\
\hline
sc (U) & 6.0326 & & 7.53 & $\begin{array}{c}220\\A=-0.02\end{array}$ & 32  & $\begin{array}{c}c_{11}=70\\c_{12}=13 \\c_{44}=-0.59 \end{array}$\\
\hline
hcp (S) & 6.4097 & 10.3624 & 0 & $\begin{array}{c}184\\\end{array}$& 14  & $\begin{array}{c}c_{11}=62\\c_{12}=42\\c_{13}=-58\\c_{33}=150\\c_{55}=24\end{array}$\\
\hline
rh (U) & 7.2232 & 88.5$^o$ & 1.62 & $\begin{array}{c}188\\\end{array}$ &  38 
&$\begin{array}{c}c_{11}=72\\c_{12}=72\\c_{13}=13\\c_{14}=-4\\c_{33}=45\\c_{44}=3\end{array}$\\
\hline
\end{tabular}
}
\caption {The same as in Table \ref{tabb} but for the SOC calculations. }
\label{tabc}
\end{center}
\end{table}
\begin{table}[th!]
\begin{center}
\resizebox{0.5\textwidth}{!} {
\begin{tabular}{c|c|c|c|c|c|c}
\hline
& $a$[a.u.] & $c$[a.u.] &  $\Delta E$[mRy/at.] & $V_{at.}$[a.u.$^3$] & $B$ [GPa] & $c_{ij}$[GPa]\\
\hline\hline
fcc (U) & 8.7214 & & 0.93  & $\begin{array}{c}166\\A=-36\end{array}$ & 63  & $\begin{array}{c}c_{11}=62 \\c_{12}=64 \\c_{44}=36 \end{array}$\\
\hline
bcc (S) & 6.9118 & & 0 & $\begin{array}{c}165\\A=9.75\end{array}$ & 62 & $\begin{array}{c}c_{11}=67 \\c_{12}=59 \\c_{44}=39 \\ \end{array}$\\
\hline
sc (S)  & 5.6120 & & 1.25 & $\begin{array}{c}177\\A=0.42\end{array}$ & 54 & $\begin{array}{c}c_{11}=63 \\c_{12}=49 \\c_{44}=3 \end{array}$\\
\hline
hcp (S) & 6.1781 & 10.0889 & 1.11 & $\begin{array}{c}167\\\end{array}$ & 45  & $\begin{array}{c}c_{11}=70\\c_{12}=56\\c_{13}=27\\c_{33}=59\\c_{55}=15\end{array}$\\
\hline
rh (U) & 6.9911 & $\alpha=85.5^{\circ}$ & 2.19 & $\begin{array}{c}169\\\end{array}$ & 56  
&$\begin{array}{c}c_{11}=45\\c_{12}=14\\c_{13}=82\\c_{14}=9\\c_{33}=40\\c_{44}=-0.02\end{array}$\\
\hline
\end{tabular}
}
\caption {The same as in Table \ref{tabc} but in addition to the SOC the relativistic local orbital for the $p_{1/2}$ state (RLO) is included.}
\label{tabd}
\end{center}
\end{table}

\end{document}